\documentclass{elsart-3}

 \usepackage{graphicx}

\usepackage{amssymb}

\usepackage[english,francais]{babel}

\newtheorem{e-proposition}[theorem]{Proposition}

\newtheorem{e-definition}[theorem]{Definition\rm}

\renewcommand{\theequation}{\arabic{equation}}
\def\og{\leavevmode\raise.3ex\hbox{$\scriptscriptstyle\langle\!\langle$~}}
\def\fg{\leavevmode\raise.3ex\hbox{~$\!\scriptscriptstyle\,\rangle\!\rangle$}}

\begin{document}

\begin{frontmatter}


\selectlanguage{english}
\title{ The Leidenfrost effect: from quasi-spherical droplets to puddles}


\selectlanguage{english}
\author[authorlabel1]{Yves Pomeau},
\ead{pomeau@lps.ens.fr}
\author[authorlabel2]{Martine Le Berre}
\ead{martine.le-berre@u-psud.fr}
\author[authorlabel3]{Franck Celestini},
\ead{Franck.Celestini@unice.fr}
\author[authorlabel4]{Thomas Frisch}
\ead{thomas.frisch@inln.cnrs.fr}

\address[authorlabel1]{Department of Mathematics, University of Arizona, Tucson, USA.}
\address[authorlabel2]{Institut des Sciences Mol\'eculaires d'Orsay ISMO-CNRS, Univ. Paris-Sud, Bat. 210, 91405 Orsay, France.}
\address[authorlabel3]{Laboratoire de Physique de la Mati\`ere Condens\'ee, CNRS UMR 7366, Universit\'e de Nice Sophia-Antipolis,
   Parc Valrose  06108 Nice Cedex 2, France}
\address[authorlabel4]{
Institut Non Lin\'eaire de Nice, CNRS UMR 7735, Universit\'e de Nice Sophia-Antipolis,
1361 Routes des lucioles, Sophia Antipolis F-06560 Valbonne France}


\medskip
\begin{center}
{\small Received *****; accepted after revision +++++\\
Presented by £££££}
\end{center}

\begin{abstract}
In the framework of the lubrication approximation, we derive a set of equations describing the steady bottom profile of Leidenfrost drops coupled with the vapor pressure. This allows to derive scaling laws for the geometry of the concave bubble encapsulated between the drop and the hot plate under it. The results agree with experimental observations in the case of droplets with radii smaller than the capillary length $R_c$ as well as in the case of puddles with radii larger than $R_c$.

{\it To cite this article:Y. Pomeau, M. Le Berre, F. Celestini, T. Frisch, C. R.
Mecanique (2012).}

\vskip 0.5\baselineskip

\selectlanguage{francais}
\noindent{\bf R\'esum\'e}
\vskip 0.5\baselineskip
\noindent
{\bf L'effet Leidenfrost: Description des profiles de gouttes et galettes }

Dans le cadre de l'approximation de lubrification l'effet Leidenfrost est d\'{e}crit par deux \'{e}quations coupl\'{e}es pour le profil  d'une goutte de Leidenfrost et
la pression de la vapeur. Ce mod\`{e}le permet de trouver les \'{e}chelles de longueur caract\'{e}risant la bulle de vapeur encapsul\'{e}e entre la goutte et la plaque chauffante. Les profils num\'{e}riques sont en bon accord avec les observations exp\'{e}rimentales, tant pour les gouttes de rayon inf\'{e}rieur \`{a} la longueur capillaire que pour les galettes de grande \'{e}tendue horizontale.

{\it Pour citer cet article~:Y. Pomeau, M. Le Berre, F. Celestini, T. Frisch,
 C. R.
Mecanique ...}

\vskip 0.5\baselineskip
\noindent{\small{\it Mots-cl\'es~:} Mot-cl\'e1~de~la~liste~; Mot-cl\'e2~;
Mot-cl\'e3}}
\end{abstract}
\end{frontmatter}


\selectlanguage{english}
\section{Introduction}
\label{intro}
\textit{It is a great pleasure to write this piece of science for Paul Clavin. Over the years he inspired us in many ways. The topic we have chosen mixes ideas of non-equilibrium science, of fluid mechanics and thermodynamics. We also predict various properties which
agree with the experimental observations. We hope that Paul will feel that this piece of science is pragmatic enough !}

The Leidenfrost effect is named after J.G. Leidenfrost (1715-1794) who wrote an article \cite{Leide}, in latin, on his observation that liquid droplets do not touch very hot surfaces and so survive much longer than normally expected. As explained by Tyndall \cite{Tyndall} on nineteenth century, the vapor released in the gap between the hot plate and the droplet lifts it  and cuts direct physical contact with the hot plate. This increases the lifetime of the evaporating droplet because of the poor heat conductivity of vapor compared to the one of the hot plate. All this works if the droplet is not too heavy. We discussed recently \cite{PRL} scaling laws for small Leidenfrost droplets. This showed the remarkable fact that, for radius less than (in order of magnitude) $R_l = \left(\frac{  \eta  \delta T \lambda}{g L \rho_v \rho_l}\right)^{1/3}$, the droplet takes off from the hot plate to reach higher and higher elevations as
the droplet gets smaller and smaller by evaporation, $(\delta T)$ being the temperature difference between the hot plate and the boiling point of the liquid, $\eta$ the shear viscosity of the vapor, $\lambda$ its heat conductivity, $g$ the acceleration of gravity, $L$ the latent heat, and $\rho_{l}$,$\rho_{v}$  the mass density of the liquid and vapor respectively. Typically  $R_l$ is in the range of a few tens micrometers ($19 \mu m$ for a drop of  water on a hot plate at $400^0 C$), that is much smaller than  the capillary length $R _c = \left(\frac{\sigma}{\rho_l g}\right)^{1/2}$, $\sigma$ being the surface tension between liquid and vapor. For water $R_c\sim 2.8 mm$, namely several order of magnitude larger than $R_l$. Therefore we shall assume $R_l \ll R_c$.

The reference \cite{PRL} dealt with very small droplets of radii of order $R_l$ or smaller, typically from $1 \mu m$ to $30 \mu m$.  In this range it was predicted and observed that
the height $h$ of the gap below the droplet increases as $R$ decreases, contrary to what is usually claimed ($h$ is also the thickness of the film of vapor between the droplet and the hot plate). Starting from droplets with radius larger than $R_l$, it was shown that, as they evaporate and when the radius becomes of order $ R_l$, then $h$ becomes of order of the horizontal gap extent $l$ and the droplets spontaneously take-off from the substrate, becoming too light to stand the upward force generated by the pressure due to evaporation. Because of this lift-up the  lubrication approximation (for the temperature field and the flow in the gap between the droplet and the hot plate) breaks down.

We focus  below on larger droplets, big enough to remain close to the hot plate, so that the lubrication approximation applies.  We derive first a set of two coupled equations for the height $h(r)$ of the droplet bottom surface and the local pressure $p(r)$. Increasing the values of $R$,  the solutions are found for the four following regimes  depending on the location of the droplet radius with respect to the radii $[R_l, R_i,R_c,\infty]$, with $R_i$ defined below.  Quasi-spherical droplets are described in section \ref{sec:undisturbed}. They exist for a radius  larger or of order $R_l$ , but much smaller than $R_i$.
described in section \ref{sec:xi1}, they correspond to radius much larger than $R_l$ , up to about $R_i$. In this case the pressure in the gap is much smaller than Laplace's pressure (we mean by Laplace's pressure the pressure drop across the vapor-liquid interface due to surface tension and equal to $\sigma(1/R_1 + 1/R_2)$, $R_{1.2}$ principal radii of curvature of the surface), therefore an "uniform approximation" can be used in the gap, and a spherical shape can be assumed close to the bottom of the droplet.
  In section \ref{sec:xigrand} we shall investigate larger droplets, of radius larger than $R_i$, up to about $R_c$. This requires a more complex study because an uniform approximation cannot be used in the gap which splits into two domains, a trapped bubble and a narrow neck connecting the bubble to the outside. Different scaling laws apply in the trapped bubble and in the neck, although the lubrication approximation remains correct in both domains.
Finally we consider in section \ref{sec:puddle} the case of puddles, with radii much larger than $R_c$.
  Note that the  fluid motion and the temperature field in the gap are well described by the lubrication approximation for all cases with $R$ larger than $R_l$.

Compared to recent publications on the same subject \cite{quere} and \cite{eggers}, this work seems to be the first one giving (original) estimates of the Leidenfrost effect as a function of the physical parameters by discussing the joint phenomena of evaporation from the droplet and the viscous vapor flow in the gap between the droplet and the hot plate. This led us to the introduction of the length scale $R_l$, which is central in our discussion, as well as the other length scale, $R_i$, depending on it and on the capillary length. To the best of our knowledge the set of equations (\ref{eq:finale1}), (\ref{eq:fz}) and (\ref{eq:press.1}), is used for the first time for solving  this problem. We notice that in another context similar looking droplet shapes have recently been described \cite{eggers}. In the latter case the drops levitate by air cushion above a porous mould through which an air stream is forced. They also display a trapped bubble related to the outside by a neck. In our work  the flow is a consequence of the evaporation and so of temperature gradient in the gap, not an imposed quantity as in  \cite{eggers}, leading to equations and solutions with scaling different to ours.

\section{Equations}
 \label{sec:eqs}

\subsection{Velocity field, pressure at the interface  and temperature field }
The lubrication approximation in the gap relies on  the three following ingredients.

{\it{i)}} Stokes equations for the flow in the gap.

This gap extends mostly in the horizontal direction $(x,y)$, and the components of the fluid velocity are $(u,v, w)$, $w$ vertical velocity. The boundary conditions are $u = v = w = 0$ for $z =0$, the Cartesian equation of the hot plate.  The other boundary conditions are on the surface of the droplet, at an elevation $z = h(x,y)$. In the lubrication limit, this surface is close to horizontal, so that the b.c. (boundary conditions) are $ u = v = 0$ for $z = h(x,y)$.  The b.c. for $w$ is Stefan condition, written as $w = \frac{\lambda T_{,z}}{L \rho_v}$
where ${T_{,z}}|_{ z = h(x,y)} $ is the derivative of the temperature with respect to $z$ on the surface of the droplet, computed on the vapor side (hereafter the notation $f_{,z}$ will be for $\frac{\partial f}{\partial z}$). Stefan condition expresses the conservation of energy: the heat flux $\lambda T_{,z}$ normal to the surface of the droplet balances the rate of transformation of liquid into vapor times the latent heat.

The Stokes equations read

\begin{equation}
\eta \nabla^2 u - p_{,x} = 0
 \mathrm{,}
\label{eq:Stokesx}
\end{equation}
\begin{equation}
\eta \nabla^2 v - p_{,y} = 0
 \mathrm{,}
\label{eq:Stokesy}
\end{equation}
and
\begin{equation}
\eta \nabla^2 w - p_{,z} = 0
 \mathrm{,}
\label{eq:Stokesz}
\end{equation}
$p$ being the pressure.
The velocity field is divergenceless, so that
\begin{equation}
 u_{,x} + v_{,y} + w_{,z} = 0 \mathrm{.}
 \label{eq:incomp}
 \end{equation}

\bigskip
\bigskip

{\it{ii)}} Balance of normal forces on the surface.

There is another equation for the shape of the liquid surface in the gap.  Let us consider the case of droplets with radius $R$ much smaller than $R_c$. Depending if the pressure generated in the gap by the evaporation flow is of order or much less than Laplace's pressure $\frac{2\sigma}{R}$ in the droplet, the equation for the surface can be discarded (section \ref{sec:undisturbed}) or not (next sections).

 If the gap pressure is much less than $\frac{2 \sigma}{R}$, one can assume that the droplet is almost spherical. This yields
\begin{equation}
 h(x,y) = h_0 + \frac{r^2}{ 2R}
 \mathrm{,}
 \label{eq:hsph}
 \end{equation}
 $h_0$ is the point on the spherical surface the closest to the hot plate and $r = \sqrt{x^2 + y^2} $ is the horizontal distance to this point. The parabolic approximation for $h(r)$ is valid in the lubrication limit, $h_0 \ll R$. It is derived in the limit $ r \ll R$ from the Cartesian equation of a circle,
 $ (R -h(r) + h_0)^2 + r^2 = R^2 \mathrm{.}$

If the fluid pressure in the gap is of the same order of magnitude as  $\frac{2 \sigma}{R}$, another equation is needed for $h(x,y)$. This equation results from the balance of normal forces on the surface of the droplet. Inside the droplet the pressure is dominated by Laplace's pressure (recall that we assume that $R$ is much smaller than the capillary radius $R_c$), whereas on the vapor side the normal stress is $ p - \eta w_{,z}$. Therefore the balance of normal forces on the surface of droplet inside the gap writes
\begin{equation}
 \frac{2\sigma}{R} - ( p - \eta w_{,z}) =  \sigma (h_{.xx} + h_{,yy})
  \mathrm{,}
 \label{eq:press}
 \end{equation}
which becomes in the axis-symmetric case
\begin{equation}
 \frac{2\sigma}{R} - ( p - \eta w_{,z}) =  \frac{\sigma}{r} \left(rh_{.r} \right)_{,r}
  \mathrm{.}
 \label{eq:pressaxiss}
 \end{equation}
Note that this condition is obviously satisfied by the quasi-spherical profile (\ref{eq:hsph}) if the pressure in the gap, $( p - \eta w_{,z})$, is negligible with respect to $\frac{2\sigma}{R}$.

{\it{iii)}} Laplace's equation for the temperature Field.

It writes $\nabla^2 T(x, y, z) = 0$ because we neglect the convective part of this flux, assuming the Peclet number to be small. This temperature field satisfies two boundary conditions: on the hot plate $ T(z=0)= T_0$ , and $ T= T_1$ on the surface of the droplet , namely for $ z = h(x, y)$. In the lubrication limit, the solution of Laplace's equation is
$$ T = T_0 \left( 1 - \frac{z}{h(x,y)} \right) + T_1 \frac{z}{h(x,y)} \mathrm{.}$$
Therefore the vertical velocity on the surface of the droplet is $w = - k \frac{\delta T}{h}$ with $\delta T = T_0 - T_1$ (a positive quantity)  and $k = \frac{\lambda}{L \rho_v}$.

\subsection{Pressure in the vapor flow}

Using the above relations, let us derive the equation for the pressure of the flow in the gap.
By integrating the incompressibility condition from $z = 0$ to $z = h(x,y)$ one obtains,
\begin{equation}
 <<u_{,x} + v_{,y}>> + w (z = h) = 0 \mathrm{.}
 \label{eq:incompint}
 \end{equation}
where $<<u>> = \int_0^{ z = h} u {\mathrm{d}}z$. In Stokes equation, $\nabla^2$ is dominated by the second derivative with respect to $z$, the shortest length scale in the lubrication limit. Therefore, $u$ is close to the Poiseuille value,
$$ u = \frac{p_{,x}}{2 \eta} z ( z - h ) \mathrm{,}$$ or  $<<u>> =  -  \frac{p_{,x} h^3}{12 \eta}$. Once   put into the equation (\ref{eq:incompint}) it gives,
\begin{equation}
\nabla_2 \cdot (\frac{h^3}{12} \nabla_2 p ) +  \frac{\eta k \delta T}{h} = 0
 \mathrm{,}
 \label{eq:finale1}
 \end{equation}
where $\nabla_2 = {\bold{e}}_x \frac{\partial}{\partial x} +  {\bold{e}}_y \frac{\partial}{\partial y}$, $\mathbf{e}_x$ being the unit vector in the $x$ direction.

The equation (\ref{eq:finale1}) is valid for all situations where the lubrication approximation applies. Given $h(x,y)$, it can be written as an Euler-Lagrange condition of minimization of the (Rayleigh) functional with respect to variations of $p$,

 \begin{equation}
{\mathcal{D}}_{Ra} = \int {\mathrm{d}}x  \int {\mathrm{d}}y \left[ \frac{h^3}{24} \left(\nabla_2 p \right)^2 - \frac{\eta k \delta T}{h} p \right]
 \mathrm{,}
 \label{eq:finale2}
 \end{equation}

For axis-symmetric geometries equation (\ref{eq:finale1}) reads explicitly,
\begin{equation}
\left(\frac{ r h^3}{12}p_{,r}\right)_{.r} +  r \frac{\eta k \delta T}{h} = 0
 \mathrm{,}
 \label{eq:finale2}
 \end{equation}
 which can be solved by a double integration,
\begin{equation}
p (r) = - (12 \eta k \delta T) \int_0^r \frac{ {\mathrm{d}} r_1}{r_1 h^3(r_1)} \int_0^{r_1} \frac{r_2 {\mathrm{d}} r_2}{ h(r_2)} + p_0 \mathrm{,}
\label{eq:pressurea}
\end{equation}
 where $p_0$ is an integration constant fixed by the boundary conditions.
 Note that this expression is valid for any $h(r)$ and requires only that the horizontal extension of the gap is much larger than its thickness.

Let us scale out the various physical quantities which have been introduced. As seen in the next section a convenient choice is

   \begin{equation}
 \left \{ \begin{array}{l}
  h_s=R_l^{3/2} R^{-1/2}\\
r_s=(R h_s)^{1/2}=R_l^{3/4}R^{1/4}\\
p_s= \rho_l g \frac{R^2}{h_s}
\mathrm{,}
\end{array}
\right. \label{eq:scale1}
\end{equation}
 as units for $h$, $r$, and $p$ . For quasi-spherical droplets, this choice readily derives from the balance between the weight of the droplet and the upward force generated by the pressure. We shall see later that it is also pertinent for the description of disturbed surfaces (droplets with radius smaller than $ R_c$), whereas another scaling will be derived for the description of Leidenfrost puddles which forms at $R >R_c$, see section \ref{sec:puddle}.

 With such scalings the equation for the flow in the gap read without any physical parameter. In the axis-symmetric case it reads,
 \begin{equation}
\left(\frac{r h^3}{12} p_{,r} \right)_{,r} + \frac{r}{h} = 0
 \mathrm{,}
 \label{eq:finale1norm}
 \end{equation}

 and
  equation (\ref{eq:pressurea}) becomes
\begin{equation}
p (r) = - 12 \int_0^r \frac{ {\mathrm{d}} r_1}{r_1 h^3(r_1)} \int_0^{r_1} \frac{r_2 {\mathrm{d}} r_2}{ h(r_2)} + p_0 \mathrm{,}
\label{eq:pressurenorm}
\end{equation}

Thanks to this integral solution, one can see that, given $h(r)$, the integration constant $p_0$ is fixed by the condition that $p$ tends to zero as $r$ tends to infinity. The constant $p_0$ is related to $h(r)$ by the expression
\begin{equation}
 p_0 = 12 \int_0^{\infty} \frac{ {\mathrm{d}} r_1}{r_1 h^3(r_1)} \int_0^{r_1} \frac{r_2 {\mathrm{d}} r_2}{ h(r_2)} \mathrm{.}
 \label{eq:pressionzero}
 \end{equation}

The balance of the vertical forces on the drop leads to an additional relation. The weight of the drop $Mg$ has to be compensated by the vertical force generated by the evaporative flow, obtained by integration of the pressure over the surface of the sphere in the gap,

\begin{equation}
 F_{z} = 2 \pi \int_0^{\infty} {\mathrm{d}} r r p(r) =Mg
\mathrm{.}
\label{eq:fz}
\end{equation}

Using the relations (\ref{eq:scale1}) the balance of vertical forces becomes in dimensionless form,
\begin{equation}
\frac{2}{3}  = \int_0^{\infty} {\mathrm{d}}r r p(r)\mathrm{.}
\label{eq:balancep}
\end{equation}

\subsection{Pressure versus curvature}
From $w_{,z} = \frac{ k \delta T}{h^2}$ one finds that $p$ is larger than the viscous stress $\eta w_{,z}$, by a factor $l^2/h^2$, $l$ horizontal extent of the gap . Therefore, when the lubrication approximation applies, the contribution of the viscous stress to the balance of vertical forces can be neglected, and equation (\ref{eq:press}) yields
\begin{equation}
 \frac{2 \sigma}{R} - p = \sigma \nabla_2^2 h
\mathrm{,}
\label{eq:press.1}
\end{equation}
which writes in
 the axis-symmetric case,
\begin{equation}
 \frac{2 \sigma}{R} - p  = \sigma(h_{,r^2} + \frac{1}{r} h_{,r}) \mathrm{.}
 \label{eq:Laplace}
 \end{equation}
After it is written with the units given in (\ref{eq:scale1}) this equation becomes
\begin{equation}
 2 - \xi p  = h_{,r^2} + \frac{1}{r} h_{,r} = \frac{1}{r} ( r h_{,r})_{,r}
 \mathrm{,}
 \label{eq:hdef}
 \end{equation}
where all the variables are scaled, making appear the
 dimensionless number
\begin{equation}
\xi = R^{7/2} R_l^ {-3/2} R_c^{-2}
\mathrm{,}
\label{eq:xi}
\end{equation}
which is the ratio of the pressure in the vapor flow to Laplace's pressure in the drop.
The scaled coupled set of  equations (\ref{eq:finale1norm})-(\ref{eq:hdef}) together with appropriate boundary conditions make up the model we shall consider henceforth. We recall that they are derived thanks to the scalings (\ref{eq:scale1}) appropriate for drops radii smaller than $R_c$, as studied in sections \ref{sec:undisturbed}-\ref{sec:xi1}-\ref{sec:xigrand}, whereas another set of scalings will be derived for larger drops (puddles), leading to the same system of dimensionless equations, see section \ref{sec:puddle}.

  \section{Solution in the case of an undisturbed surface}
 \label{sec:undisturbed}

 In this section we solve the equations for the lubrication limit in the case where the droplet remains almost spherical, that implies to neglect the term $\xi p$ in equation (\ref{eq:hdef}). This eliminates the need to derive the shape of the surface of the droplet in the gap by using the balance of normal forces. The present case (undisturbed sphere) is a fairly standard application of lubrication theory. We start using the original variables and show that the scalings proposed in (\ref{eq:scale1}) naturally yield parameterless equations.

  The result of the integration of the right hand side of equation (\ref{eq:pressurea}) with $h(r)$ given by equation (\ref{eq:hsph}), is
 $$ p(r) = p_0 -  \frac{12 \eta k R \delta T}{4 h_0^3} G(\frac{r^2}{h_0R})\mathrm{,}$$ where $ G (\alpha)$ is the numerical function defined as $$ G (\alpha) = \int_0^{\alpha} \frac{{\mathrm{d}} \alpha'\ln(1 + \alpha'/2)}{\alpha'(1 + \alpha'/2)^{3}} \mathrm{.}$$

To have a pressure tending to zero at $r$ tending to infinity one must take $p_0 =  \frac{12 \eta k R \delta T}{4 h_0^3} G (\infty)$, whence the result,
\begin{equation}
p(r) = \frac{12 \eta k R \delta T}{4 h_0^3}  \left( G(\infty) - G(\frac{r^2}{h_0 R})\right)
 \mathrm{,}
 \label{eq:finale3}
 \end{equation}
which yields the vertical force generated by the evaporative flow, $F_z$ defined in (\ref{eq:fz}),

$$ F_{z}= \pi \frac{3 \eta k R^2 \delta T}{8 h_0^2}  \mathrm{.}$$

By writing that this force balances exactly the weight of the sphere,
\begin{equation}
2 \pi \int_0^{\infty} {\mathrm{d}} r r p(r) = M g
\mathrm{,}
\label{eq:balancev}
\end{equation}

one finds $h_0 = \sqrt{\frac{3 \pi}{8} \frac{R_l^3} {R}}$,
that agrees with the scaling proposed in (\ref{eq:scale1}) for the  gap height.
Whenever the surface of the droplet  is close to a parabolic cylinder, equation (\ref{eq:hsph}) for $h(r)$ in scaled variables  becomes

\begin{equation}
h(r)  = h_0 + \frac{r^2}{2} \mathrm{. }
\label{eq:hnorm}
\end{equation}

Notice that $p_0$ is now a pure number constrained by the condition that $p(r)$ tends to zero as $r$ tends to infinity, and that $h_0$  in equation (\ref{eq:hnorm}) is also a pure number
defined by the balance of vertical forces on the droplet (\ref{eq:balancep}).

  The set of equations solve the quasi-spherical problem if the lubrication approximation applies, namely if the height $h_0$ is physically much less than $R$, which requires $R \gg R_l$.
  It also assumes that the pressure in the gap is negligible compared to $\sigma/R$, because we assumed the relation (\ref{eq:hsph}). This requires $\xi \ll 1$, or $R \ll R_i$ with
 \begin{equation}
R_i = (R_l^3 R_c^4)^{1/7}
\label{eq:Ri}
\end{equation}

which is about $330\mu m$ for water over a plate heated at $400^0$C . Otherwise, one has to determine the shape of the droplet in the gap, namely the function $h(r)$, as done next.

   \section{Solution in the case of a disturbed surface}
 \label{sec:xi1}

In this case the surface of the droplet in the gap is not a spherical cap because the pressure generated by fluid motion there is not negligible with respect to
Laplace's pressure inside the droplet. Compared to the previous case, we have to solve the same equation for $p$ already written in (\ref{eq:finale1}), but the profile being unknown, we  need also to solve equation (\ref{eq:press}) with the convenient boundary conditions.  This second equation is derived from the balance of normal forces. Assuming the equation (\ref{eq:finale1}) solved, the pressure $p$ is known.

 In the range $\xi \sim 1$ the new equation to be considered is the equation for the curvature of the droplet in the gap, which should replace the simple relation (\ref{eq:hsph}) used in section \ref{sec:undisturbed} for the range $\xi \ll 1$ characterizing an undisturbed spherical droplet. Finally,
if $\xi$ is not small, equations (\ref{eq:finale1}) and (\ref{eq:press.1}), or their scaled form (\ref{eq:finale1norm}) and (\ref{eq:hdef}), make together a pair of equations allowing to obtain  the  droplet profile $h(r)$ and the pressure $p(r)$ in the gap.

 Some properties can be derived without explicitly solving the equations. From the integral solution for the pressure given in (\ref{eq:pressurenorm}), the pressure is a decreasing function of $r$ because $h(r)$ is positive. Because it has to tend to zero at $r$ infinite, the pressure is positive and decreasing. From equation (\ref{eq:hdef}) the mean curvature $ \frac{1}{r} ( r h_{,r})_{,r} $ is an increasing function of $r$, tending to $2$ (in dimensionless units) as $r$ tends to infinity. This excludes in particular very large values of this curvature at finite values of $r$ in the limit $\xi$ large.

 The condition of balance of vertical forces can be transformed into a condition for the behavior of $h(r)$ at large $r$. In the dimensionless version of the equations, this condition is derived by integrating both sides of equation  (\ref{eq:hdef}) from zero to a large radius with the element of integration $(r \mathrm{d}r)$. The final result yields the following condition valid at $r$ tending to infinity:
 \begin{equation}
h(r) \approx \frac{r^2}{2} - \frac{2 \xi}{3} \ln(r) +...
 \mathrm{.}
 \label{eq:hintegr}
 \end{equation}
 This is a way of expressing that the vertical force on the droplet is equal to the uncompensated vertical component of the capillary forces.

\begin{figure}[htbp]
\centerline{$\;\;$
(a)\includegraphics[height=1.75in]{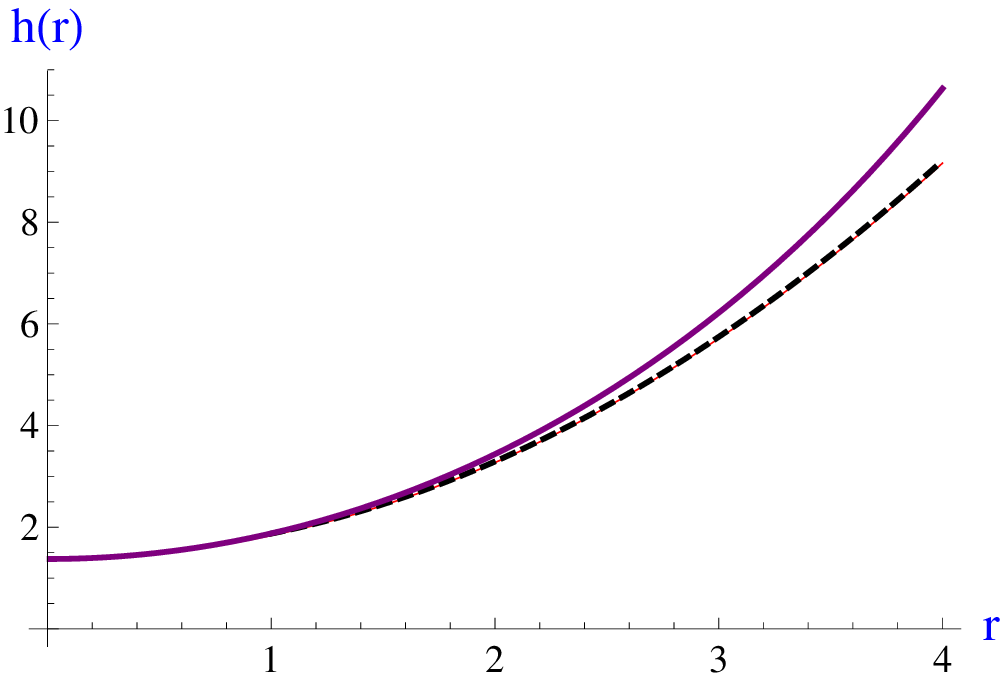}
(b)\includegraphics[height=1.5in]{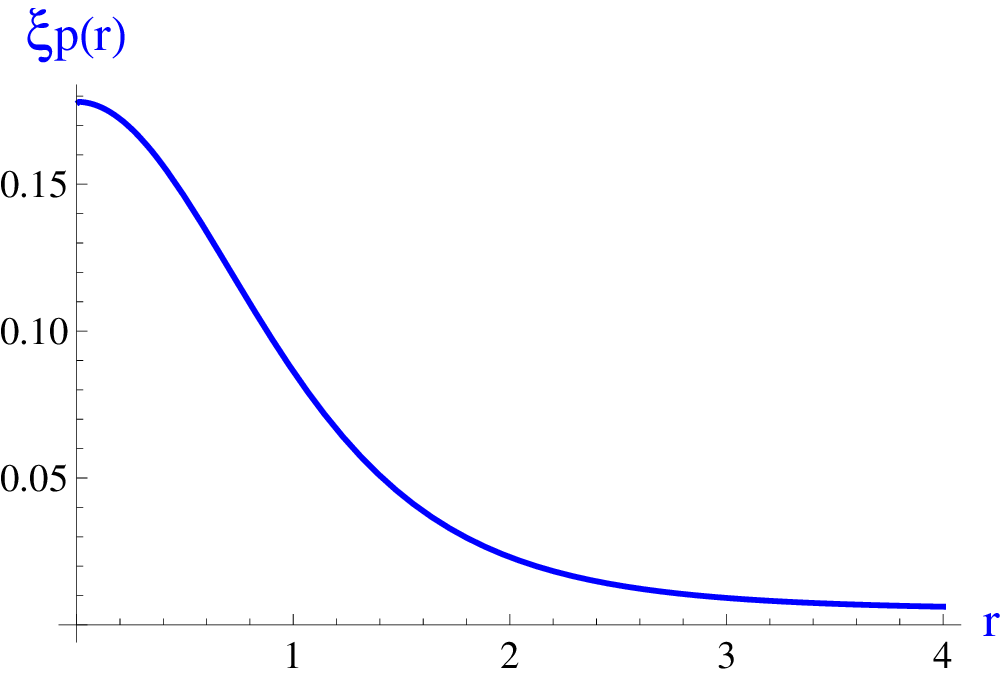}
 $\;\;$}
\caption{
(a) Bottom profile  of  a Leidenfrost drop for $\xi=0.18$ or $R=200\mu m$. The numerical solution of equation (\ref{eq:hfinal}) in red solid line melting with the asymptotic expansion (\ref{eq:hintegr}) in black dashed line, are located below the spherical  profile (solid purple line). Initial conditions $h_{,r^2}(0)=(2-\xi)/2$, $h_0=1.375$. The physical variables can be recovered by using the scaling lengths $h_s=5.86\mu m$, $r_s=34\mu m$.
(b) Decreasing pressure versus axial distance.
$\xi p(r)$ solution of equations (\ref{eq:hdef}).
}
\label{Fig:xipetit}
\end{figure}

 One can reduce the equations to a single one for $h(r)$ with a closed set of b.c. By simple algebra, one derives:
  \begin{equation}
- \frac{1}{\xi} \left(\frac{r h^3}{12}\left( h_{,r^2} + \frac{1}{r} h_{,r}\right)_{,r}\right)_{,r} + \frac{r}{h} = 0  \mathrm{.}
 \label{eq:hfinal}
 \end{equation}
The Laurent expansion of $h(r)$ near $r = 0$ reads $h(r) = h_0 + a_1 r^2 + a_2 \ln(r) + a_3 r^3+...$ where $h_0$ and $a_{1-3}$ are free coefficients, although the coefficients of the next order terms in the expansion, like $b r^4$, etc. can be derived order by order from equation (\ref{eq:hfinal}).  The coefficients $a_2$ and $a_3$ must vanish to make the solution smooth. The two remaining free parameters $h_0$ and $a_1$ are fixed by the asymptotic behavior of the solution at large $r$.
 It reads $$ h(r) = c r^2 + d \ln(r) + \frac{f}{r^2} + ...\mathrm{,}$$ where $c$, $d$, and $f$ are free parameters. Two parameters are constrained by the condition that  $a_2$ and $a_3$ vanish, but the solution that we shall display has $a_2 = a_3 = 0$ and so we shall not consider them anymore, and call $a_1$ simply $a$. The next order terms are also derivable from $c$ and $d$ by order by order solution of the equation (\ref{eq:hfinal}). The b.c. on the shape of the surface imposes $c = 1/2$ and the condition for the balance of vertical forces imposes $ d = - \frac{2 \xi}{3 }$. This yields two conditions for two free parameters, $h_0$ and $a$.
The exact solution of equations (\ref{eq:hfinal}) fits well  the analytical expansions (\ref{eq:hintegr}) and (\ref{eq:hnorm}) (valid for $r<1$ and $r>1$ respectively), as shown in Fig.\ref{Fig:xipetit} where the solid line coincides with the dotted line, as distinct from the spherical profile located above.

\section{Limit $\xi$ large}
 \label{sec:xigrand}

The limit $\xi$ small is somewhat trivial since in this limit one can neglect the term $\xi p$ in equation (\ref{eq:hdef}) which has the simple solution $h(r) = h_0 + \frac{r^2}{2}$, and one is back to the case of a spherical droplet of unit radius, as expected.

An obviously interesting limit is the limit of a large $\xi$.
In this limit, the solution splits into two different domains. Those domains are derived from an analysis of the solution of the equations in this limit of a large $\xi$. The results presented are consistent with the equations, although they cannot be considered as obvious consequences of them. Between $r = 0$ and $r = r_c$ (to be found) the surface of the sphere is like a trapped bubble with a negative curvature and where the pressure $p$ is almost constant. The radius $r_c$ is such that the trapped bubble solution crosses the hot surface, which is obviously impossible.  To get rid of this crossing, other scalings must be used locally and a neck replaces this crossing, as shown below. From equation (\ref{eq:hdef}) the mean curvature of the surface is also constant, and negative (which is the only possible choice as one can check).

\subsection{Unbalanced pressure hypothesis}

 Assuming for the moment (something that will be shown not to be correct) that this constant value is not $2/\xi$, the curvature of the surface of the trapped bubble should of order 1, one finds that $h_b (r)$, namely the value of $h(r)$ in the trapped bubble is equal to
$$h_b (r) = \frac{2 - \xi p_0}{4} r^2 + h_0
 \mathrm{.}$$
We assume now that the main contribution to the upward force is from the bubble, something to be checked at the end. This yields from equation (\ref{eq:hnorm}) that $$r_c^2 =  \frac{4 h_0}{p_0 \xi - 2} \mathrm{.}$$
This gives a trapped bubble which is an almost perfect spherical cap with constant pressure inside. It should be connected to the outside by a neck smoothing the solution near $r = r_c$, as necessary because the integral contribution to the right-hand side of equation (\ref {eq:pressurenorm}) scales formally like $r^2/h^4$, of order $\xi^{-3}$ in the bubble although $p_0$ scales like $\xi^{-1}$, which makes it dominant. Therefore inside the trapped bubble the pressure is constant and equal to $p_0$.  The balance of vertical forces yields
$$ h_0 = \frac{p_0 \xi - 2}{3 \pi p_0} \mathrm{.}$$ This solution does not work however because it cannot be matched with the solution in the neck. In other words we have to discard the hypothesis of spherical cap (for the surface of the trapped bubble), because the continuity of slope of the surface at the transition between the neck and the trapped bubble cannot be insured.

\subsection{Description of the trapped vapor bubble}

The only possibility remaining is that the pressure inside the bubble is at leading order (with respect to $\xi$) equal to $2/\xi$ to balance Laplace's pressure inside the droplet, plus a small contribution depending on $r$ and balancing the curvature of the droplet there. Therefore we assume that, inside the trapped bubble, $$ p = \frac{2}{\xi} + p_b \mathrm{,}$$ where $p_b (r)$ is much smaller than $\xi^{-1}$. At leading order $ p = \frac{2}{\xi} $. This makes it straightforward to derive the radius of the trapped bubble from equation (\ref{eq:balancep}) which set out the balance of vertical forces. One finds
\begin{equation}
r_c^2  = \frac{2\xi}{3}
 \mathrm{.}
\label{eq:radiusbub}
\end{equation}

This radius is actually the radial distance between the axis at $r =0$ and the neck where the pressure makes a transition from its value $2/\xi$ inside the bubble to zero outside of it. Therefore the radius inside the trapped bubble is of order $\xi^{1/2}$, the order of magnitude of $r_c$. The equation relating $p_b$ to the curvature of the surface reads
 \begin{equation}
- \xi p_b  = h_{b,r^2} + \frac{1}{r} h_{b,r}
 \mathrm{.}
\label{eq:press1bub}
\end{equation}
The other equation relating $h_b$ and $p_b$ is derived from (\ref{eq:finale1norm}) and reads
 \begin{equation}
\left(\frac{r h_b^3}{12} p_{,r} \right)_{,r} +  \frac{r}{h_b} = 0
 \mathrm{.}
\label{eq:press2bub}
\end{equation}
Defining scaled quantities $\xi^{-1/2}r$,  $\xi^{-3/5} h_b$ and  $\xi^{+7/5} p_b$ makes disappear any small or large parameter in the differential equations to be satisfied by those quantities. It means also that, inside the trapped bubble, the height is of order $\xi^{3/5}$ and that the correction to the leading order constant pressure ($2/\xi$ in the original variables) is of order $\xi^{-7/5}$, negligible in the large $\xi$ limit, as expected, with respect to the leading order contribution $2/\xi$. Note also that even though the scaling law $h_b \sim \xi^{3/5}$ seems to imply that the lubrication approximation does not hold because $h$ seems to be much bigger than $r_c  \sim \xi^{1/2}$, this is not so because physically $h$ and $r$ are  originally measured with different unit lengthes. We shall come back to this at the end of this section.

With the scaled quantities (written the same as the original quantities) the equations to be satisfied are equation (\ref{eq:press2bub}) with $p_b$ instead of $p$ and
 \begin{equation}
-  p_b  = h_{b,r^2} + \frac{1}{r} h_{b,r}
 \mathrm{.}
\label{eq:press1bub.1}
\end{equation}
 The condition to be satisfied by this set of equations is $h_b (r_c) = 0$ with $r_c =  \left(\frac{2}{3}\right)^{1/2}$.
Actually this crossing is unphysical. It defines the large distance behavior (in inner variables) of a neck solution connecting the trapped bubble with the outside.

\subsection{Neck region}

Let $\delta = r - r_c$ be the local coordinate in the neck and $h_n(\delta)$ be the local height. In the neck the pressure is of order $1/\xi$ because it has to tend to $p_0 = 2/\xi$ on one side (in the trapped bubble) and to zero outside. From equation (\ref{eq:balancep}) because $p$ is of order $\xi^{-1}$,  $h_n(r)$ should scale like $\delta^2$. Assuming the neck to be much less extended than the trapped bubble, namely that $\delta \ll r_c $, one finds that, inside the neck $h$ is much smaller than in the bubble, namely much smaller than $\xi^{3/5}$, and
$$ h_{n,\delta^2} = 2 - \xi p(\delta) \mathrm{.}$$
This is to be completed by equation (\ref{eq:finale1norm}) relating the pressure and the height inside the neck. As far as the order of magnitude with respect to $\xi$ is concerned this last equation is consistent  if $p$ scales like $1/\xi$, $\delta$ scales like $\xi^{1/6}$ and $h_n(\delta)$ scales like $\xi^{1/3}$ (This power law $\xi^{1/3}$ makes, as expected, the height in the neck much smaller than the height of the trapped bubble the latter being of order $\xi^{3/5}$).
Let us introduce local (overlined) quantities by absorbing the scaling laws in multiplicative factors,

   \begin{equation}
 \left \{ \begin{array}{l}
 \overline{p} = \xi p\\
\overline{\delta} = \xi^{ -1/6} \delta\\
\overline{h}_n = \xi^{-1/3} h_n
\mathrm{.}
\end{array}
\right. \label{eq:scaleneck}
\end{equation}
The equations to be satisfied by the overlined quantities are purely numerical (namely without large or small parameter) and read
\begin{equation}
\left(\frac{\overline{h}_n^3}{12}\overline{p}_{,\overline{\delta}} \right)_{,\overline{\delta}} + \frac{1}{\overline{h}_n} = 0
 \mathrm{,}
\label{eq:neck1}
\end{equation}
and
\begin{equation}
2 - \overline{p}  = \overline{h}_{n,\overline{\delta}^2}
 \mathrm{.}
\label{eq:neck2}
\end{equation}

The boundary conditions for the pressure are simple to write. For $\overline{\delta}$ very large negative (that is on the side of the bubble), the scaled pressure should tend to $\overline{p}_0 = 2 $, and as $\overline{\delta}$ tends to plus infinity (namely outside the bubble) the pressure should tend to zero. The asymptotic conditions for $h(\delta)$ are dealt with later.

Supposing $\overline{h}_n$ known, one can integrate equation (\ref{eq:neck1}) to obtain:

\begin{equation}
 \overline{p}(\overline{\delta}) = - 12 \int_{-\infty}^{ \overline{\delta}} \frac{{\mathrm{d}} \overline{\delta}_1}{\overline{h}_{n}^3 (\overline{\delta}_1)}  \int_{-\infty}^{\overline{\delta}_1}\frac{{\mathrm{d}}\overline{\delta}_2}{ \overline{h}_{n}(\overline{\delta}_2)} +  \overline{p}_0 \mathrm{,}
\label{eq:pressurenormblayer}
\end{equation}
The integration limit $r = 0$ of the original problem is pulled to $\overline{\delta} = - \infty$ for the overlined variable. The jump of pressure from $p = p_0 = 2/\xi$ to zero takes place almost exclusively across the neck. This yields $\overline{p}_0 = 2$ as a constant of integration. The other b.c. for the pressure is $\overline{p} \rightarrow 0$ as  $\overline{\delta}$ tends to plus infinity. It becomes the following condition for $\overline{h}_n$:
\begin{equation}
2 =  \overline{p}_0 = 12 \int_{-\infty}^{+\infty} \frac{{\mathrm{d}} \overline{\delta}_1}{\overline{h}_{n}^3 (\overline{\delta}_1)}  \int_{-\infty}^{\overline{\delta}_1}\frac{{\mathrm{d}}\overline{\delta}_2}{ \overline{h}_{n}(\overline{\delta}_2)}  \mathrm{.}
\label{eq:pressurenormblayer}
\end{equation}
 The solution in the trapped bubble and in the neck should merge somewhere. This merging occurs in a region where the solution of the equations on either side have a common power law behavior with respect to the distance to $r_c$. This power law behavior is easier to see on the neck side. It has to do with the way the solution behaves for $\overline{\delta}$ large negative. The equations to be solved are (\ref{eq:neck1}) and  (\ref{eq:neck2}). Let us introduce $\overline{p}_1 = 2 - \overline{p}$. The equations to be solved become

 $$ \left(\frac{\overline{h}_n^3}{12}\overline{p}_{1,\overline{\delta}} \right)_{,\overline{\delta}} =  \frac{1}{\overline{h}_n}  \mathrm{,}$$ and $$\overline{p}_1  = \overline{h}_{n,\overline{\delta}^2} \mathrm{.}$$

 They have as an exact solution the power law,
 $$ \overline{h}_n = \left(\frac{5^4}{2}\right)^{1/5} (- \overline{\delta})^{4/5} \mathrm{,}$$ and
  $$\overline{p}_1 = (- \overline{\delta})^{-6/5} \mathrm{,}$$
  that makes an acceptable solution because, at large negative $\overline{\delta}$, $\overline{p}_1$ decays to zero. Therefore the pressure tends to 2, as it should.

  Outside the neck, the solution should merge with the asymptotic spherical droplet in a region where the pressure tends to zero. For the neck solution this writes
  $\bar{h}=\bar{\delta}^2 + \alpha \bar{\delta}^k$. Introducing this expression in equations (\ref{eq:neck1})-(\ref{eq:neck2}), we obtain the asymptotic expansion valid for  large $\bar{\delta}$
   \begin{equation}
 \bar{h}_n \approx \bar{\delta}^2 + \frac{1}{120} \bar{\delta}^{-4} +...
   \label{eq:out1}
\end{equation}
   and
      \begin{equation}
 \bar{p} \approx \frac{1}{6} \bar{\delta}^{-6}
  \mathrm{.}
   \label{eq:out1}
\end{equation}
These asymptotic expansions display a surface evolving towards a sphere, and a pressure  decaying to zero at large positive $\bar{\delta}$ as expected.
  \begin{figure}[htbp]
\centerline{$\;\;$
\includegraphics[height=3in]{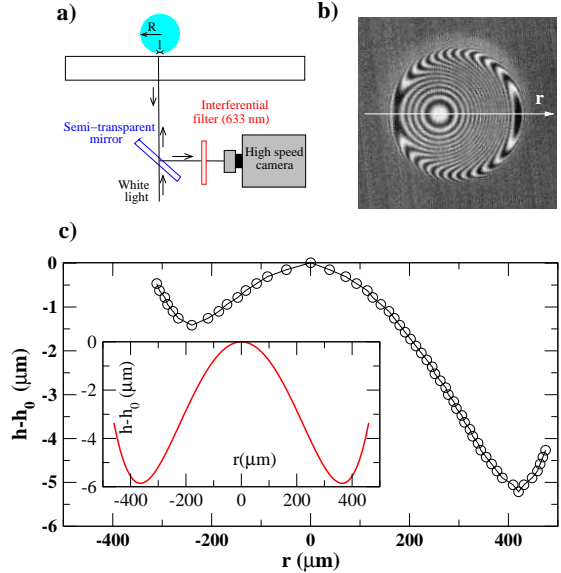}
$\;\;$}
\caption{Experiment and theory for $\xi=80$.
(a) experimental setup ; (b)  2D plot of the portion of surface below the drop (interference pattern) for water Leidenfrost drops of radius $R=1.14mm$ on a substrate kept at  $T_0=400^0 C$; (c) experimental profile $h-h_0$ (circles on the black curve) of the droplet bottom along the horizontal axis of curve (b). The red curve inserted below is the theoretical solution of  axis-symmetrical equations (\ref{eq:finale1norm})-(\ref{eq:hdef}) for  $\xi=80$, with initial conditions $h_{,r^2}(0)=-\xi^{-2/5}$, $h_0=0.9 \xi^{3/5}$  plotted in physical variables using the scalings $h_s=2.44 \mu m$, $r_s=53\mu m$, $p_s/\sigma=0.07\mu m^{-1}$ defined in (\ref{eq:scale1})
.
}
\label{Fig:expfig}
\end{figure}

The above description of a vapor bubble encapsulated below the concave part of the droplet in the large $\xi$ regime was done under the hypothesis of axis-symmetric profile, although experiments report mostly non axis-symmetric ones \cite{Celestini}-\cite{chicago}. Therefore a quantitative comparison between
numerical data and experiments is out of our scope. Nevertheless we have found a fair agreement between numerical solutions of equations (\ref{eq:finale1norm})-(\ref{eq:hdef}) and the experimental profiles for large $\xi$, as illustrated in Fig. \ref{Fig:expfig}. In this figure
the experimental set-up is depicted in (a)(see the detail in \cite{Celestini}). A high speed camera is used
to image the interference fringes between the drop-vapor and substrate-vapor interfaces which are visible in (b). The relative height profile below the drop along the horizontal of figure (b) is reported in (c), where the experimental asymmetrical black curve ( with circles) qualitatively agrees with the axis-symmetrical solution of equations (\ref{eq:finale1norm})-(\ref{eq:hdef}) obtained for the same value of parameter $\xi$.

	  \begin{figure}[htbp]
\centerline{$\;\;$
\includegraphics[height=1.5in]{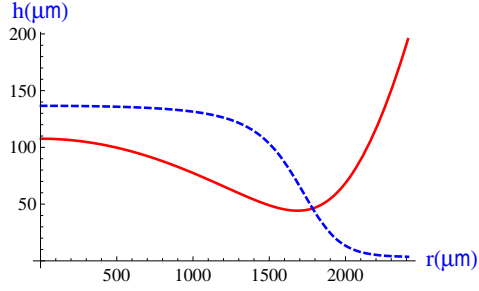}
$\;\;$}
\caption{
Numerical solution of equations (\ref{eq:finale1norm})-(\ref{eq:hdef}) for $\xi=1000$, or $R=2.36mm$. The height
 $h$ (red solid curve )  and pressure $ap$ (blue dashed curve)  are plotted in physical variables, with $a=140$ a scale factor. Initial conditions $h_{,r^2}(0)=-0.15$, $h_0=\xi^{3/5}$.
}
\label{Fig:xi1000}
\end{figure}

As shown in Figures \ref{Fig:expfig}(c) the trapped bubble size and height are clearly comparable in the numerics and in the experiment. Furthermore the trapped bubble connects with the outside through a very narrow neck, as predicted above. The narrow neck is observed for large $\xi$ values corresponding to the domain $R_i< R< R_c$ as illustrated for $\xi=1000$ in Fig.\ref{Fig:xi1000}. In the latter figure we also report the pressure profile in the gap (dashed curve) which decreases from the center of the bubble to the outside of the neck, where it vanishes.
The theoretical curves are drawn with the physical variables by scaling back the variables $h,p$ help to the expressions written in the next subsection.

  \subsection{Large $\xi$ domain in physical variables}
   \label{sec:physvar}

  The case $\xi$ large considered here is such that the droplet radius fulfills the condition $R_i \ll R \ll R_c$.
  The length scale $h_s$ for the height of the droplet above the heated plate, and the length scale for the horizontal distances $r_s$ are given by the relations \ref{eq:scale1}. Therefore the radial extent of the trapped bubble  is
    \begin{equation}
    r_b  \sim R ^{2}( R_c)^{-1}
     \mathrm{,}
     \label{eq:rb}
\end{equation}
 which  is also the radius of the disc of contact at equilibrium of small nonwetting droplets. The vertical thickness of the trapped bubble is

     \begin{equation}
   h_b  \sim R ^{8/5}R_l^{3/5}( R_c)^{-6/5}
    \mathrm{,}
     \label{eq:hb}
\end{equation}

  The  gap elongates in the horizontal direction as $R$ increases, with an aspect ratio  $h_b/r_b= R^{-2/5}R_l^{3/5}R_c^{-1/5}$  becoming smaller and smaller as $R$ increases.
 Therefore the lubrication approximation remains correct as $\xi$ gets bigger and bigger, which is equivalent to increase $R$ at constant $R_c$ and $R_l$. Therefore the range of applicability of the theory at large $\xi$ relying on the lubrication approximation extends all over the range $[R_i, R_c]$ and stops to be valid when $R$ becomes of the same order of magnitude as the capillary radius.

 \section{Evaporation of droplets in the various regimes}
 \label{sec:evap}
An obvious application of the ideas presented before is the derivation of the lifetime of the droplet. This is related to the evaporation flux from the droplet. This flux depends itself on the range of parameters where the radius lies. At very small radii one expects that, because the droplet is well above the hot surface, the evaporation flux is very close to the one for an isolated droplet (without the hot plate) in a vapor at temperature $T + \delta T$ at large distances. In the case of an isolated droplet the total 
mass evaporation rate is given in order of magnitude by
$J_{is} \sim R^2 \rho_v w $ with $w=\frac{\lambda}{L\rho_v }T_{,z}$ and $T_{,z}\sim T/R$, that gives
  \begin{equation}
J_{is} = \frac{ R\lambda \delta T}{ L}
 \mathrm{.}
    \label{eq:jis}
\end{equation}
This mass loss rate is the one explaining the decay of the square radius proportional to time, the well-known $D^2$ law.
 We shall consider below the mass loss of matter out of the droplet in various range of radii. The corresponding law of decay of the radius is easy to get in each case and we shall not do the calculation. Actually we shall limit ourselves to find whether the main contribution to the evaporative mass comes from the quasi spherical part of the droplet or from the evaporation in the gap.

 Let us compare to $J_{is}$ the evaporative mass in the vapor film between the droplet and the hot plate in the regime $R_l\ll R \ll R_i$. In this regime, the area of the film is much less than $4 \pi R^2$, the area of the drop, but the temperature gradient in the film is much larger than $\delta T/R$, so the normal speed of evaporation is much bigger than what it is for an isolated droplet with the same $\delta T$. Therefore it is not obvious which mass loss rate, $J_{is}$ or $J_{film}$ is the largest. The total mass loss rate in the film  is of order $ l^2 \rho_v w \sim  l^2 \frac{\lambda \delta T}{L h}$. From the relation $ l^2 = R h$ one derives immediately that $J_{film} \sim J_{is}$ where $J_{is}$ is given in (\ref{eq:jis}), an unexpected result valid in the regime $R_l\ll R \ll R_i$.

\begin{figure}[htbp]
\centerline{$\;\;$
\includegraphics[height=2.5in]{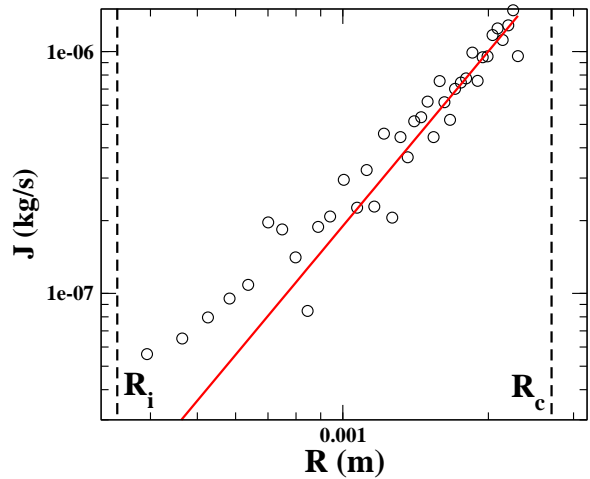}
$\;\;$}
\caption{The experimental mass evaporation rate (points) are plotted together with the theoretical prediction (solid curve). The red curve is the best fit to equation (\ref{eq:jfilm}), very close to $J_{film} =\frac{2 \lambda \delta T}{L}  \left(\frac{R^{12}}{R_c^4 R_l^3}\right)^{1/5}$.
}
\label{Fig:evapfranck}
\end{figure}

 Let us estimate $J_{film}$ in the regime $R_i\ll R \ll R_c$. We use again the relation $J_{film} \sim  r_b^2 \rho_v w \sim  r_b^2 \frac{\lambda \delta T}{L  h_b}$ together with the estimates  $r_b$ (radius of the film) and $h_b$ (thickness of the film or of the trapped bubble), given by equations (\ref{eq:rb})-(\ref{eq:hb}). Inserting those estimated into the expression of $J_{film}$ pertinent for this case, we obtain
   \begin{equation}
   J_{film} \sim \frac{ \lambda \delta T}{L}  \left(\frac{R^{12}}{R_c^4 R_l^3}\right)^{1/5} =  J_{is} \left(\frac{R}{R_i}\right)^{7/5}
   \mathrm{.}
    \label{eq:jfilm}
\end{equation}
  Therefore, in the range $R_i \ll R \ll R_c$ the ratio  $J_{film} / J_{is}$ is of order $ \left(\frac{R}{R_i}\right)^{7/5} $, much bigger than one, so that the flux is dominated by the contribution from the film. We did not consider  the flux coming from the neck domain. It is rather straightforward to show that it is negligible compared to $J_{film}$ in the limit $\xi$ large and $r_b \ll R$, equivalent to  $R_i \ll R \ll R_c$. Our theoretical prediction is confirmed by the experiment, as shown in Fig.\ref{Fig:evapfranck}. In this figure the experimental data (circles) are obtained from droplets evaporating over a thin brass substrate (thin enough to be curved and then stabilize the droplet) healed at $300^0 C$. The red curve is the best fit to equation (\ref{eq:jfilm}) in the whole domain $R_i<R<R_c$, leading to a mass evaporating rate very close to $ \frac{2 \lambda \delta T}{L}  \left(\frac{R^{12}}{R_c^4 R_l^3}\right)^{1/5}$, an expression that could be derived from $J_{film} \sim \pi r_b^2 \rho_v w $ by using the relation (\ref{eq:radiusbub}) for $r_c=r_b/r_s$.

\bigskip
\bigskip

 From this discussion one can conclude that, even though the low heat conductivity of the vapor lowers the rate of mass loss of the droplet, it does it in different ways depending of the range of droplet radius one considers.

A droplet beginning with a radius of order $R_c$ will decay to a (much smaller) radius $R_i$ after a time of order $t^* \sim (R_c^3 R_i^7)^{1/5} \frac{L \rho_l}{\lambda \rho_v \delta T}$. If one substitutes $R^2$ for $(R^3 R_i^7)^{1/5}$ in this estimate one gets back the standard $D^2$ law. This is consistent with the acceleration of the evaporation: to evaporate an isolated droplet of radius $R_c$ would take a time of order $R_c^2 \frac{L \rho_l}{\lambda \rho_v \delta T}$, much longer than the time $t^*$ just estimated (recall that $R_c \gg R_i$).

\section{Leidenfrost puddles}
 \label{sec:puddle}
 At equilibrium, drops of radius larger than $R_c$ standing on a flat plate become flat puddles. Of course they cannot be characterized anymore by their radius only, because their shape is not spherical. The relevant quantity to characterize them is their volume. If this volume $V$ were spherical, it would define a radius $R = \left(\frac{3V}{4 \pi}\right)^{1/3}$. Differently from the drop case above, we consider below physical variables, in the limit $R \gg R_c$, with $R$ so defined. In this limit, we assume that the equilibrium puddle of liquid has a circular shape of radius of order $r_c$ and thickness of order $R_c$. From the estimate of the volume of the puddle, the radius $r_c  =  \left(\frac{R^3}{R_c}\right)^{1/2}$  can be considered as given and much bigger than $R_c$.

\subsection{description of the vapor layer}
 We consider now the situation of a flat puddle hovering on a hot plate by the Leidenfrost effect, and derive the structure and thickness of the vapor layer between the plate and the puddle.  Like in the case of droplets of radius in the range $[R_i, R_c]$ we assume the puddle to be very close to its equilibrium shape at the given volume. This has to be checked at the end to yield a solution consistent with the assumptions and with the underlying physics. A direct consequence of the assumption of closeness to the equilibrium shape is that the pressure in the puddle near the bottom is almost constant and just equal to Archimedes hydrostatic value $ \rho_l g H$, $H$ being the height of the puddle, of order $R_c$, so that we shall simply replace $H$ by $R_c$ in the coming order of magnitude estimates. Therefore the situation is similar to the one studied before in the sense that the pressure in the liquid (above the vapor film) is constant, the Archimedes's pressure replacing the Laplace's one. Let $p_0 \sim \rho_l g R_c$  be this Archimedes pressure.

 The equation (\ref{eq:Laplace}) for the pressure reads now
 \begin{equation}
p_0 - p  = \sigma(h_{,r^2} + \frac{1}{r} h_{,r}) \mathrm{,}
 \label{eq:Laplacepuddle}
 \end{equation}
As we did before we assume that $p$ is the constant $p_0$ plus a small (with respect to a parameter to be found) part $p_1$ which balances the curvature term in equation (\ref{eq:Laplacepuddle}). Therefore
 \begin{equation}
- p_1  = \sigma(h_{,r^2} + \frac{1}{r} h_{,r}) \mathrm{,}
 \label{eq:Laplacepuddle.1}
 \end{equation}
This is to be completed by the equation (\ref{eq:finale2}) for the inhomogeneous part $p_1(r)$ of the pressure
\begin{equation}
\left(\frac{ r h^3}{12}p_{1,r}\right)_{.r} +  r \frac{\eta k \delta T}{h} = 0
 \mathrm{,}
 \label{eq:finale2.1}
 \end{equation}
The two equations can be reduced to a dimensionless form by taking
  \begin{equation}
 \left \{ \begin{array}{l}
 r_c  = \left(\frac{R^3}{R_c}\right)^{1/2}\\
  h_c=\left(R_l^{3} r_c^{4} R_c^{-2}\right)^{1/5}=\left(R_l^3R^6R_c^{-4}\right)^{1/5}\\
p_{1c}=(R_l^{3} r_c^{-6} R_c^{-2})^{1/5}
\mathrm{,}
\end{array}
\right. \label{eq:scale2}
\end{equation}

 as unit for $r$,  $h$ and for $p_1/\sigma$. The two resulting equations read
 \begin{equation}
- p_1  = (h_{,r^2} + \frac{1}{r} h_{,r}) \mathrm{,}
 \label{eq:Laplacepuddlessdim}
 \end{equation}
and
\begin{equation}
\left(\frac{ r h^3}{12}p_{1,r}\right)_{.r} +  r \frac{1}{h} = 0
 \mathrm{.}
 \label{eq:finale2.1}
 \end{equation}

 The condition $p_1 \ll p_0$ is satisfied if $ r_c \gg (R_c R_l)^{1/2}$, which is satisfied because we assumed $r_c \gg R_c  \gg R_l$.

There is another constraint on this solution: the film of vapor must be a thin layer underneath the puddle. Therefore its height must be much smaller than the height of the puddle. This implies $R_c \gg h$, equivalent to $ R_c \gg R_l^{3/5} r_c^{4/5} R_c^{-2/5}$, or to $R_c^7 \gg R_l^3 r_c^4$. This puts an upper bound on the radius of the puddle : $\frac{r_c}{R_c} \ll \left(\frac{R_c}{R_l}\right)^{3/4}$, which is compatible with the conditions $R_c \ll r_c$ and  $R_l \ll R_c$. Lastly a neck makes the transition between the trapped bubble and the outside.
 \begin{figure}[htbp]
\centerline{$\;\;$
(a)\includegraphics[height=1.5in]{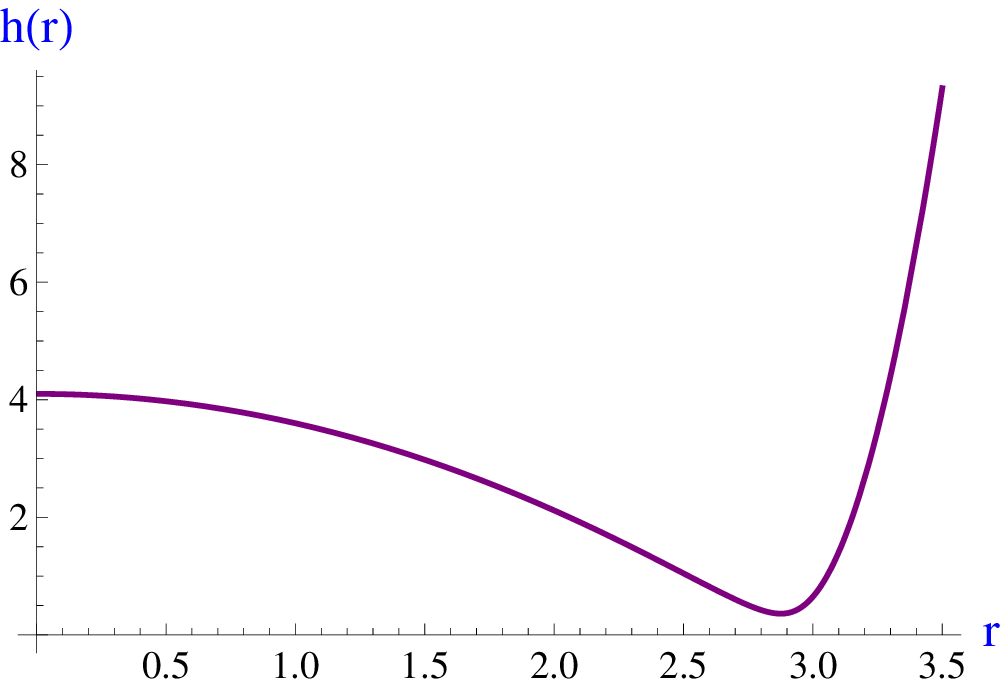}
(b)\includegraphics[height=1.5in]{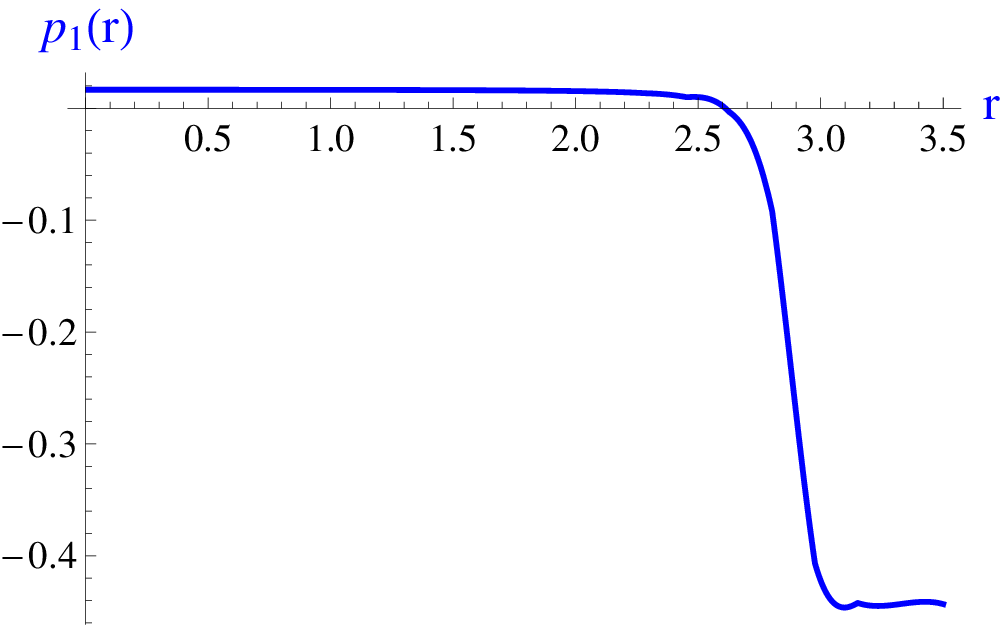}
$\;\;$}
\centerline{$\;\;$
(c)\includegraphics[height=1.5in]{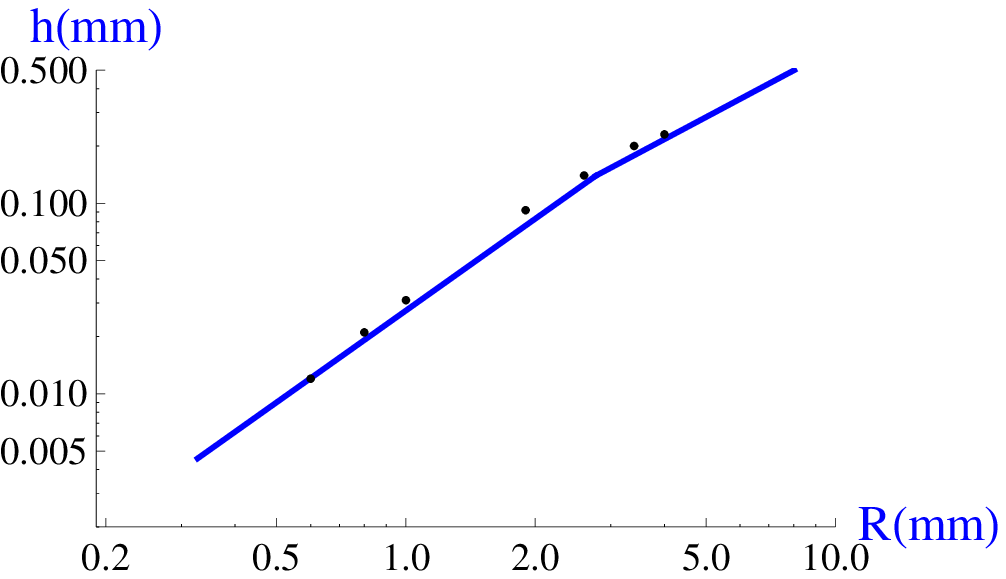}
(d)\includegraphics[height=1.5in]{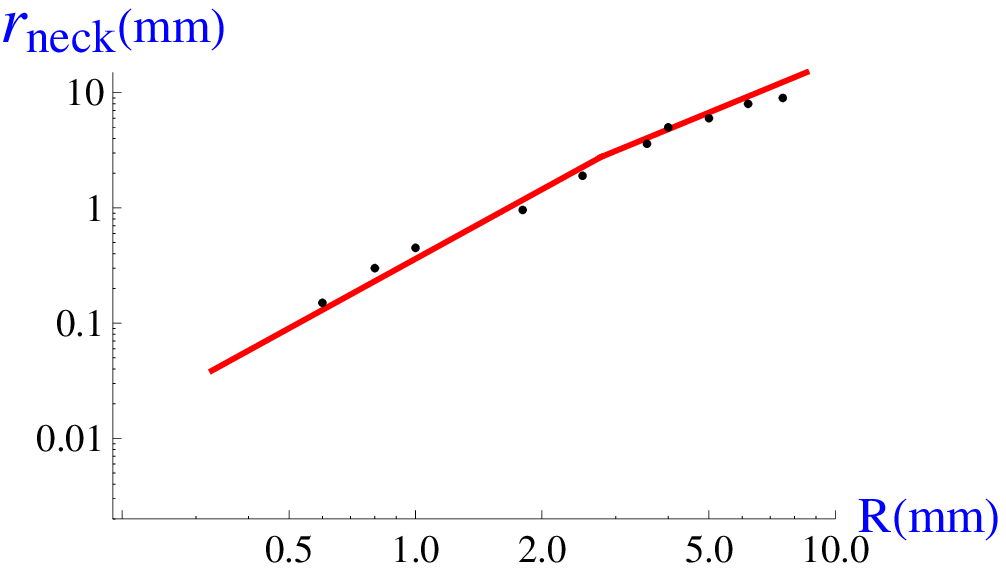}
$\;\;$}
\caption{
Solution (a) $h(r)$ and (b) $p_1/p_0$ of equations (\ref{eq:Laplacepuddlessdim})-(\ref{eq:finale2.1}) for $R=7.5 mm$. Numerical initial conditions $h_{,r^2}(0)=-1$, $h_0=4.1$ in scaled variables.
(c)-(d) Experimental (points) and theoretical (solid lines) values, of the height $h$ in (c) and neck radius $r_{neck}$ in (d) versus $R$, of the vapor bubble below the drop. The piecewise theoretical curves are plotted in physical variables, from equations (\ref{eq:rb})-(\ref{eq:hb}) for $ R_i<R<R_c$, and  relations (\ref{eq:scale2}) for $R_c<R<R_{max}$, both without any fit.
}
\label{Fig:flaque}
\end{figure}
In the neck the equations to be satisfied are formally the same as equations (\ref{eq:neck1}) and (\ref{eq:neck2}), except that the scaling are derived differently. The constant pressure in the trapped bubble is now $p_o = \rho_l g R_c$. One obtains the same equations as (\ref{eq:neck1}) and (\ref{eq:neck2}) by taking as units of width, height and pressure,
   \begin{equation}
 \left \{ \begin{array}{l}
 \overline{\delta} = \left(R_cR_l\right)^{1/2}\\
 \overline{h}=R_l\\
\overline{p}=\frac{p_0}{2}
\mathrm{.}
\end{array}
\right. \label{eq:scaleneck}
\end{equation}

The b.c. are the same as before, namely $\overline{p}$ tends to 2 as $\overline{\delta}$ tends to minus infinity and $\overline{p}$ tends to 0 as $\overline{\delta}$ tends to plus infinity.

It is remarkable that this puddle solution is given by the solution of the same set of equations, but with different scaling parameters. Remember that it has a limited range of existence because, physically, the wider is the puddle, the larger is the pressure in the trapped bubble, so that it reaches the top of the puddle at the limit of its domain of existence. It could be that at still larger masses of fluid, steady solutions have chimneys of vapor crossing the puddle from bottom to top. In reality it is likely that this corresponds to the onset of boiling, an unsteady phenomenon in general. Moreover the trapped bubble could be Rayleigh-Taylor unstable as soon as $r_c$ gets bigger than few $R_c$, an instability that could be counteracted by viscosity of the vapor in the gap. Therefore steady solutions in this range of values of $r_c\gg R_c$ are at best only indirectly connected to real life phenomena.
An example of numerical solution for the puddle is given in Figure \ref{Fig:flaque} where curves (a)-(b) display the profile and pressure for $R=R_{max}$, the largest radius of stable water puddles. We plot in curves (c)-(d), solid lines, the  theoretical values of the height and neck radius for the two domains $[R_i, R_c]$  and $[R_c,R_{max}]$  as derived in paragraphs (\ref{sec:xigrand}) and (\ref{sec:puddle}) respectively. The points reported on these figures are experimental results corresponding to Fig.2 of \cite{chicago}, that are in very good agreement with our theoretical predictions drawn without any fit to experimental data.  Let us precise that while the variable $r_{max}$ in \cite{chicago}  equals $R$ for $R<<R_c$, it is close to $R(\frac{2R}{3R_c})^{1/2}$ for $R\gg R_c$. In between we have extrapolated the $r_{max}(R)$ dependance in order to compare the data of \cite{chicago} with our theoretical results.
\subsection{evaporation rate}
The mass evaporation rate of the vapor layer below puddles can be derived by using similar argument as used for droplets in section \ref{sec:evap}. It writes $J_{film} \sim  r_c^2 \rho_v w \sim  r_c^2 \frac{\lambda \delta T}{L  h_c}$ where $r_c$ and $h_c$ are the the radius and height of the film. Introducing  the physical values written in equation (\ref{eq:scale2}), we obtain
\begin{equation}
   J_{film} \sim \frac{ \lambda \delta T}{L}  \left(\frac{R^{9}}{R_c R_l^3}\right)^{1/5} =  J_{is} \left(\frac{R}{R_i}\right)^{4/5}\left(\frac{R_c}{R_i}\right)^{3/5}
   \mathrm{,}
    \label{eq:jfilm2}
\end{equation}
which is still much larger than the evaporation rate of the equivalent spherical isolated drop. Let us compare this
 mass loss rate (out of the vapor layer) with the rate out of the top of the puddle, and also with the rate out of the neck region. To calculate the temperature gradient close to the top surface of the puddle, we see the puddle as a disc of radius $r_c$ and temperature $T-\delta T$ inserted on a plate heated at temperature $T$. This leads to a vertical temperature gradient of order $\delta T/r_c$ because $r_c$ is the unique scale length of this problem. It follows that the evaporation rate is $J_{top}\sim  r_c^2 \frac{\lambda \delta T}{L  r_c} \sim \frac{h_c}{r_c}J_{film}$ at the top of the puddle. On the other hand the mass loss rate in the neck region is $J_{neck}\sim  r_c  \overline{\delta}\frac{\lambda \delta T}{L   \overline{h}} \sim  \frac{ \overline{\delta}}{r_c }\frac{h_c}{\overline{h}}J_{film}$.  Using equations (\ref{eq:scale2})-(\ref{eq:scaleneck}), these two relations lead to the ratios
\begin{equation}
  \frac{ J_{top}}{J_{film}} \sim \left(\frac{R_l}{\sqrt{RR_c}} \right)^{3/5}
   \mathrm{,}
    \label{eq:jtop}
\end{equation}

and
\begin{equation}
  \frac{ J_{neck}}{J_{film}} \sim \left(\frac{R_lR_c^2}{R^3}\right)^{1/10}
   \mathrm{,}
    \label{eq:jneck}
\end{equation}
that are much smaller than unity for $R_l\ll R_c< R \leq R_{max}$ ( recall that $R_{max}$ is of order few $R_c$). We conclude that the evaporation mostly takes place in the vapor layer  below the puddle, as it was shown for droplets in the range $R_i< R <R_c$ in section \ref{sec:evap}. In summary the evaporation process happens through the vapor bubble for any drop or puddle whose radius $R$ belongs to the domain $[R_i, R_{max}]$.

\section{Summary and perspective.}
This contribution explains the scaling laws for the Leidenfrost phenomenon, sweeping the domain of small droplets to large puddles. Our approach relies on a scaling analysis of the fundamental equations in the lubrication approximation, which has a wide domain of applicability for explaining the Leidenfrost phenomenon. A significant restriction on the validity of our approach is a constraint on the control parameters: we assumed $R_c \gg R_l$, in agreement with the data for the experiments done in Nice. It could be however that this inequality is not satisfied in other experiments, opening the way to other scaling laws. Another very interesting question is the transition to boiling. It has to do with the extension of our approach to large puddles. Experimentally this could be related to the much studied and still mysterious effect of film boiling. There one has to deal with the occurrence of physical contact between the liquid and the hot plate by the breaking of the continuity of the vapor film, something beyond our approach, as we assumed the hot plate to be at fixed temperature, above the boiling temperature of the liquid. Likely the understanding of the transition to film boiling requires a solution of the heat transfer equations in the solid also. The ratio of heat conductivities of the vapor and the solid could be used as a small parameter for this problem.

Another limit for the applicability to real experiments of the concepts of fluid mechanics and heat transfer, as used in the present work,  is the Knudsen limit: the mean-free path of molecules in ordinary conditions in air is in the micrometric range, and could be of the same order or even bigger than some length scales of the Leidenfrost  phenomenon. Indeed the smallest scale is the thickness of the film of vapor underneath the droplet. If this thickness gets noticeably smaller than the mean free path, the correct physical picture for the flow there is by the direct solution of the equations of kinetic theory, the Boltzmann equation in principle or simplified versions of it, like the BGK model \cite{bgk}. In the limit of a mean-free path much bigger than the thickness of the vapor layer, one recovers a rather simple description: the molecules bounce on both sides of the thin layer and so make a Brownian motion in the horizontal direction. This is described in the lubrication approximation by a diffusion equation with the time and the horizontal coordinates  as variables, and the number density per unit horizontal area as conserved quantity. This has to be matched with the regular continuum mechanics picture (Stokes equation and Laplace's equation for the temperature field). This transition to a rarefied gas situation could explain some of the observation of splashing of droplets at low pressure \cite{nagel}. In this much studied problem of the impact of droplets on solid surfaces (in the absence of any temperature effect), the pressure in the trapped bubble should depend on the history of the collision, and some scaling laws derived in this paper could be valid. For example small impacting droplets at low speed could remain quasi spherical and so could be dealt with the same lubrication approximation that we used. At larger impact velocities it is likely that a concave trapped bubble could show up too. All this will be  the subject of future investigations.

\section*{Acknowledgements}
We greatly acknowledge  J.C. Burton et al. who sent us their experimental data files.

      \end{document}